\documentstyle[12pt]{article}
\begin{document}
\baselineskip=24pt

\title{\begin{Huge}
Quantum Black Holes\\
As Elementary Particles\\
\end{Huge}
\vspace{1.2in} }
\author{Yuan K. Ha\\ Department of Physics, Temple University\\
 Philadelphia, Pennsylvania 19122 U.S.A. \\
 yuanha@temple.edu \\    \vspace{2in}  }
\date{December 30, 2008}  
\maketitle
\newpage
\begin{center}
\begin{large}
{\bf Abstract}
\end{large}
\end{center}
\vspace{.3in} 
Are black holes elementary particles? Are they fermions or bosons?
We investigate the remarkable possibility that quantum black
holes are the smallest and heaviest elementary particles. We are able to
construct various fundamental quantum black holes: the spin-0, spin-1/2,
spin-1, and the Planck-charge cases, using the results in general relativity. 
Quantum black holes in the neighborhood of the Galaxy could resolve
the paradox posed by the Greisen-Zatsepin-Kuzmin limit on the energy of 
cosmic rays from distant sources. They could also play a role
as dark matter in cosmology.\\


\newpage
An elementary particle is a logical construction.
We would like to investigate the possibility that quantum black holes are the smallest and heaviest elementary particles [1]. 
A puzzling situation arises when one extends the results of a
classical black hole to the Planck scale region. If a black hole
has an asymptotic mass equal to the Planck mass $M_{Pl} = \sqrt{\hbar c/G}$,
then its Schwarzschild radius is found to be twice the Planck length
$l_{Pl} = \sqrt{\hbar G/c^{3}}$, i.e.
\begin{equation}
R_{S} = \frac{2GM_{Pl}}{c^{2}} = 2 l_{Pl} .
\end{equation}
Here $\hbar$ is Planck's constant, $G$ is the gravitational constant and 
$c$ is the speed of light.
Thus the Planck length is inside the horizon of the black hole and is never accessible.
This is a bizarre situation. In order to probe the physics at the
Planck length, an agent must be able to reach down to the Planck length and to
depart from the Planck length in principle.
A new theorem for black holes has been established and which can remove the above difficulty. 
{\em The horizon mass theorem states that for all black holes: 
neutral, charged or rotating, the horizon mass is always equal to twice the irreducible mass
observed at infinity} [2,3]. If the horizon mass of a black hole is taken to be the Planck mass $M_{Pl}$, 
then a distant observer will find the black hole to have an asymptotic mass $M_{Pl}/2$ and
therefore its Schwarzschild radius is exactly equal to the Planck length, i.e.
$R_{S} = l_{Pl}$. Since the diameter of the black hole is $2R_{S}$, this length becomes
identical to the Compton wavelength $\lambda$ of the black hole as determined by the distant
observer using the asymptotic mass $M$, i.e.
\begin{equation}
\lambda = \frac{\hbar}{Mc} = \frac{\hbar}{(M_{Pl}/2)c} = 2R_{S}.
\end{equation}
We therefore consider a spin-0 quantum black hole to have a horizon mass $M_{Pl}$ and a
Schwarzschild radius $l_{Pl}$. It is one of the smallest and heaviest
conceivable elementary particles.\\

A spin-0 quantum black hole is by definition a boson. It could be
created in ultra-high energy collisions or in the Big Bang, but
it will disintegrate immediately after it is formed and become
Hawking radiation. The observable signature of such a quantum
black hole can be seen from its radiation. Ultra-high energy cosmic rays corresponding to
extraordinary energies beyond $10^{20}$ eV have been observed but their source locations have
not been identified [4,5,6]. Quantum black holes in the neighborhood of our Galaxy could
resolve the paradox posed by the Greisen-Zatsepin-Kuzmin limit [7] on the energy of cosmic rays
from distant sources. As we shall see, quantum black holes carrying maximal charges are
absolutely stable but they can annihilate with opposite ones to produce bursts of
elementary particles. These could be the ultra-high energy cosmic ray events which have been
detected but not yet explained in several observations.\\

{\em A quantum black hole can be stable}.
Consider a black hole with charge $Q$ and asymptotic mass $M$. An
exact energy expression of such a Reissner-Nordstr\"{o}m black hole
has also been found [2].
The total energy contained within a radius at coordinate $r$ outside the black hole is given by
\begin{equation}
E(r) = \frac{rc^{4}}{G} \left[ 1 - \sqrt{1 - \frac{2GM}{rc^{2}} + \frac{GQ^{2}}{r^{2}c^{4}} }\right] .
\end{equation}
In the extreme charged case, the following condition holds: $GM^{2} = Q^{2}$ .
Analysis of the energy distribution in Eq.(3) shows that the electrostatic energy of a charged black hole resides completely outside the black hole and in the extreme case this electrostatic energy is exactly cancelled by the negative gravitational energy there. The net energy outside the black hole is identically zero. 
As a result the horizon mass is seen to be the same as the asymptotic mass by a
distant observer.  The black hole is therefore absolutely stable and it cannot emit any radiation.\\

A Reissner-Nordstr\"{o}m black hole has a radius given by 
\begin{equation}
r_{+} = \frac{GM}{c^{2}} +  \frac{GM}{c^{2}} \sqrt{1 - \frac{Q^{2}}{GM^{2}} },  
\end{equation}
and the extreme horizon radius is $R_{H} = GM/c^{2}$.
If the horizon mass of such an extreme black hole is taken to be the Planck mass $M_{Pl}$, 
then its asymptotic mass is still $M_{Pl}$. The extreme horizon radius
is again the Planck length $l_{Pl}$, which is its Compton wavelength
$\lambda = {\hbar/M_{Pl}c}$ according to the distant observer. 
This is the Planck-charge $Q_{Pl}$ quantum black hole.\\

A remarkable consequence follows from the properties of these
extreme charged quantum black holes. For two similar Planck-charge
quantum black holes separated at any distance, their electrostatic
repulsion exactly cancels their gravitational attraction so that
they become non-interacting particles. There is no effective
potential between them at all. Since the Planck length is
extraordinary small, of the order of $10^{-33}$ cm, any
particle-size distance such as $10^{-16}$ cm is huge compared to
that length. The condition of asymptotic flatness for these
quantum black holes is quickly reached even at the very short
distance of $10^{-30}$ cm. They effectively move in a flat
spacetime and behave like classical point masses obeying the laws
of classical mechanics. Each constituent carries one unit of
Planck mass $2.176 \times 10^{-5}$ gm and one unit of Planck length
$1.616 \times 10^{-33}$ cm as radius. A collection of these quantum
black holes in a finite volume is a non-interacting gas. We
recognize that they are the most natural candidate for dark matter
in galaxies without having to resort to new interactions and exotic particles.\\

Two opposite Planck-charge quantum black holes could form a bound state until they annihilate
each other with a powerful burst of radiation. First, they would combine into a neutral black hole with their charges neutralized. The resulting black hole would then evaporate via Hawking radiation. The radiation would consist of a large number of elementary particles in all directions with very high energies, thereby providing an observable signature of these quantum black holes.\\

{\em A quantum black hole can possess a spin}. It may be considered as a microscopic rotating black hole,
using the Kerr solution. The total energy of a Kerr black hole for all rotations can be calculated in the teleparallel equivalent of general relativity by numerical evaluation [8]. 
An approximate analytical energy expression for a slowly rotating black hole is given by [9] 
\begin{eqnarray}
E(r) & = & \frac{rc^{4}}{G} \left[ 1 - \sqrt{ 1 - \frac{2GM}{rc^{2}} + \frac{a^{2}}{r^{2}}  }  \right]            \nonumber \\
     &   &  + \frac{a^{2}c^{4}}{6rG} \left[ 2 + \frac{2GM}{rc^{2}} + \left(1 + \frac{2GM}{rc^{2}} \right)
              \sqrt{ 1 - \frac{2GM}{rc^{2}} + \frac{a^{2}}{r^{2}}  }  \right],
\end{eqnarray}
where $a = J/Mc$ is the angular momentum parameter. $M$ is the mass of the black hole including rotational energy and $J$ is the angular momentum. Analysis of the rotating black hole in the numerical evaluation shows that for {\em all values} of the angular momentum parameter $0 \leq a \leq GM/c^{2}$ the horizon mass is almost perfectly twice the irreducible mass. The tiny discrepancy from the exact value of two is due to evaluating the rotating black hole with axial symmetry by a configuration with spherical symmetry. The rotational energy of a Kerr black hole again resides completely outside the black hole as the electrostatic energy does in the charged black hole case.\\

A Kerr black hole with angular momentum $J$ has a radius given by
\begin{equation}
r_{+} = \frac{GM}{c^{2}} +  \frac{GM}{c^{2}} \sqrt{1 - \frac{J^{2}c^{2}}{G^{2}M^{4}} }. 
\end{equation}
For maximal angular momentum, $J_{max} = GM^{2}/c$,
the horizon radius from Eq.(6) becomes $R_{H} = GM/{c^{2}}$ .
By requiring this radius to be the minimum Planck length $l_{Pl}$, we see that the Kerr mass is the same as
the Planck mass $M_{Pl}$. Therefore the maximum angular momentum for such a quantum black hole is
$J = GM_{Pl}^{2}/c = \hbar$ 
according to a distant observer. The Compton wavelength, $\lambda = {\hbar/M_{Pl}c}$, is
again the Planck length $l_{Pl}$. We consider this to be a spin-1
quantum black hole which is a boson.\\

When a maximally rotating black hole is observed to have an asymptotic mass $M$, its irreducible mass,
corresponding to the mass seen by a distant observer if the rotation is removed by the Penrose process, 
is $M_{irr} = M/\sqrt{2}$ [10]. According to the horizon mass theorem stated earlier, 
the horizon mass of the extreme Kerr black hole is $2M_{irr} = \sqrt{2} M$. 
Such a black hole, however, is not stable against Hawking emission since its horizon mass is greater than its asymptotic mass. Consequently, the spin-1 quantum black hole has a horizon mass $\sqrt{2} M_{Pl}$ and 
it will decay into a burst of elementary particles.\\

A Kerr black hole with $J = \hbar$ so constructed excludes the possibility of it being electrically charged by the condition for a general charged rotating black hole with electric charge $Q$, (in the c.g.s. units, the Couloumb constant is unity)
\begin{equation}
\left( \frac{GM}{c} \right)^{2} \geq G \left( \frac{Q}{c} \right)^{2} + \left( \frac{J}{M} \right)^{2}
\end{equation}
because the maximum angular momentum $J$ of the black hole has saturated the inequality.\\

{\em A quantum black hole can possess both a spin and a charge}.
A Kerr-Newman black hole with charge $Q$ and angular momentum $J$ has a radius given by
\begin{equation}
r_{+} = \frac{GM}{c^{2}} +  \frac{GM}{c^{2}} \sqrt{1 - \frac{Q^{2}}{GM^{2}} - \frac{J^{2}c^{2}}{G^{2}M^{4}} }. 
\end{equation}
The maximal charge and angular momentum combination corresponds to the equality sign in Eq.(7), in which
case the horizon radius also becomes  $R_{H} = GM/{c^{2}}$ . Again by requiring this radius to be the minimum Planck length $l_{Pl}$, we find that the Kerr-Newman mass has to be same as the Planck mass $M_{Pl}$. 
The Compton wavelength is then $\lambda = l_{Pl}$, exactly the same as the horizon radius. 
A remarkable possibility is to consider $J = \hbar/2$ and $Q = \sqrt3 Q_{Pl}/2$ in Eq.(8), 
where the Planck charge is defined by $Q_{Pl}^{2} = GM_{Pl}^{2}$.  
This would result in a spin-1/2 quantum black hole carrying an electric charge. 
A spin-1/2 quantum black hole is by definition a fermion according to a distant observer in an asymptotically flat spacetime.\\

A spin-1/2 quantum black hole is also unstable against Hawking radiation as it will decay into
a burst of elementary particles. The irreducible mass in this case is found to be 
$M_{irr} = \sqrt 5 M_{Pl}/4$, while the horizon mass is $M_{H} = \sqrt 5 M_{Pl}/2$ by the use of the horizon mass theorem. The horizon mass is therefore greater than the asymptotic mass 
$M_{Pl}$. A spin-1/2 quantum black hole further has a magnetic moment equal to $\sqrt3 Q_{Pl}l_{Pl}/4$.\\

We are thus able to construct several fundamental quantum black holes: the spin-0, spin-1/2, spin-1 
and the Planck-charge $Q_{Pl}$ quantum black holes.
Each quantum black hole has the defining property that it has exactly the minimum Planck length as its radius. If we extend Hawking's result on black hole entropy [11] to a spin-0 quantum black hole
then we may obtain the smallest unit of entropy in quantum gravity,
\begin{equation}
S = \frac{1}{4} \left( \frac{kc^{3}}{\hbar G} \right) A
  = \pi k \left(\frac{A}{4\pi l^{2}_{Pl}} \right) = \pi k ,
\end{equation}
where $k$ is the Boltzmann constant. If the quantum black hole is
an elementary particle, then such a particle is seen to possess an
intrinsic gravitational entropy. This intrinsic entropy is
presumably a measure of its disorder during its creation. An extreme charged quantum black
hole cannot evaporate and therefore its entropy is locked, unless it is
annihilated by an oppositely charged quantum black hole.
This situation may explain the paradox of information loss in gravitational collapse 
with the possible production of numerous stable quantum black holes in the process.\\

Quantum black holes have a microscopic size but a macroscopic mass.
{\em More significantly, their Compton wavelength is equal to their classical size.}
If the Compton wavelength of an object is larger than its classical size,
then a quantum description of the object is necessary. This is the case 
of the electron. The Compton wavelength of an electron is about $10^{-10}$ cm,
while its classical radius is about $10^{-13}$ cm. Thus a classical theory 
of the electron is inadequate. On the other hand, if the classical radius 
of an object is larger than its Compton wavelength, then a classical description
is sufficient. A 1-kg meteorite has a typical size of $10^{1}$ cm but its Compton wavelength
is only about $10^{-41}$ cm. Obviously there is no need for a quantum gravity theory
of the meteorite even though its mass is several orders of magnitude higher than
the Planck mass. Quantum black holes are at the very boundary between classical and
quantum regions. They obey the Laws of Thermodynamics and they decay into
other particles. Therefore a semiclassical description of them is appropriate.
Quantum black holes also provide a natural cutoff to 
spacetime at the Planck scale and they could render general relativity 
in the weak field perturbative expansion directly finite. 
There is no need to calculate Feynman integrals over
arbitrarily small volumes and to deal with the renormalizability
issues as in the conventional field theory approach. 
They may also explain the origin of ultra-high energy cosmic rays and dark matter in
cosmology. They could have a real existence and their discovery would be indispensable
to understanding the ultimate nature of spacetime and gravity.\\

\end{document}